\documentclass[12pt]{elsarticle}

\usepackage{graphicx}
\graphicspath{{images/}}

\usepackage{amssymb}

\newcommand{\bea}{\begin{eqnarray}}
\newcommand{\eea}{\end{eqnarray}}
\newcommand{\be}{\begin{equation}}
\newcommand{\ee}{\end{equation}}

\usepackage{color}
\usepackage[normalem]{ulem}
 
\journal{Journal of Colloid and Interface Science}

\begin{document}

\begin{frontmatter}



\title{Why do magnetic nanoparticles form messy clumps? Taking into account the bridging or sticking of ligands in simulations}


\author[add1]{N.~R.~Anderson}
\author[add2]{D.~R.~Louie\corref{cor1}}
\author[add1,add3,add4]{D.~Serantes}
\author[add1,add2]{K.~L.~Livesey}
\cortext[cor1]{Present address: Department of Astronomy, University of Maryland, College Park, MD 20742}

\address[add1]{UCCS Biofrontiers Center, University of Colorado - Colorado Springs, 1420 Austin Bluffs Parkway, Colorado Springs CO 80918, USA}
\address[add2]{Department of Physics, University of Colorado - Colorado Springs, 1420 Austin Bluffs Parkway, Colorado Springs CO 80918, USA}
\address[add3]{Instituto de Investigaci\'ons Tecnol\'oxicas and Applied Physics Department, Universidade de Santiago de Compostela, 15782 Santiago de Compostela, Spain}
\address[add4]{Department of Physics, University of York, York YO10 5DD, UK}

\begin{abstract}
Experiments on magnetic nanoparticles in a viscous medium have shown that agglomerates form that display complex shapes. However, most theoretical results predict more simple, ordered shapes, such as single-particle width chains. To account for this discrepancy we have created a theoretical model that phenomenologically includes the bridging or ``stickiness" between ligands on nearby nanoparticles.
This interaction is accounted for through a unitless stickiness parameter $c$ that can be varied between 0 (no bridging between ligands) and 1 (irreversible bridging or sticking together on impact). An analytic estimate for the value of $c$ is provided based on a comparison between the time for a particle to diffuse to an agglomerate compared to the time for it to reorient into the local magnetic field direction. Numerical Langevin simulations are performed using ferromagnetic, 50~nm and 25~nm diameter magnetite nanoparticles with various ligand coating lengths in hexane in order to support the analytic estimate. The simulations produce agglomerates that are qualitatively and quantitatively similar to experimental results.
\end{abstract}


\end{frontmatter}


\section{Introduction}
\label{intro}

Ferromagnetic and superparamagnetic nanoparticles in liquids are of great interest for a variety of applications including guided drug delivery \cite{DobsonReview}, magnetic hyperthermia (heat) therapy \cite{Fortin}, magnetically-tuned liquid crystal displays \cite{WangLC}, lab on a chip applications \cite{LeeLab}, and MRI contrast agents \cite{Bonnemain}. When the particles are in fluids, there is the possibility for them to agglomerate due to the long-range magnetic dipole-dipole interactions.

Ferromagnetic nanoparticles will form loops to minimize the magnetic dipole-dipole energy in zero field, as long as the magnetic interaction is stronger than both thermal fluctuations \cite{Wei2011,Jund95} and isotropic electrostatic interactions. In a moderate applied magnetic field, they will instead form single-particle-width chains.\cite{Cheng} These two scenarios are drawn in Fig.~1 in panels (a) and (b). Although this behavior is predicted by many simulations\cite{rovigatti,Wang02}, in experiments clumps of these particles are far more compact and are more ``messy" than simulations predict, as illustrated by the cartoon in Fig.~1$($c$)$.\cite{Saville:2013hf,martinez} Sometimes rather than chains, clumps which are almost isotropic in shape form.\cite{saville2014formation} Moreover, real agglomerates are often far thicker than those produced through simulation. The question therefore is: why are the experimental clumps so messy? 

\begin{figure}[h]
\centering
\includegraphics[width=0.7\textwidth]{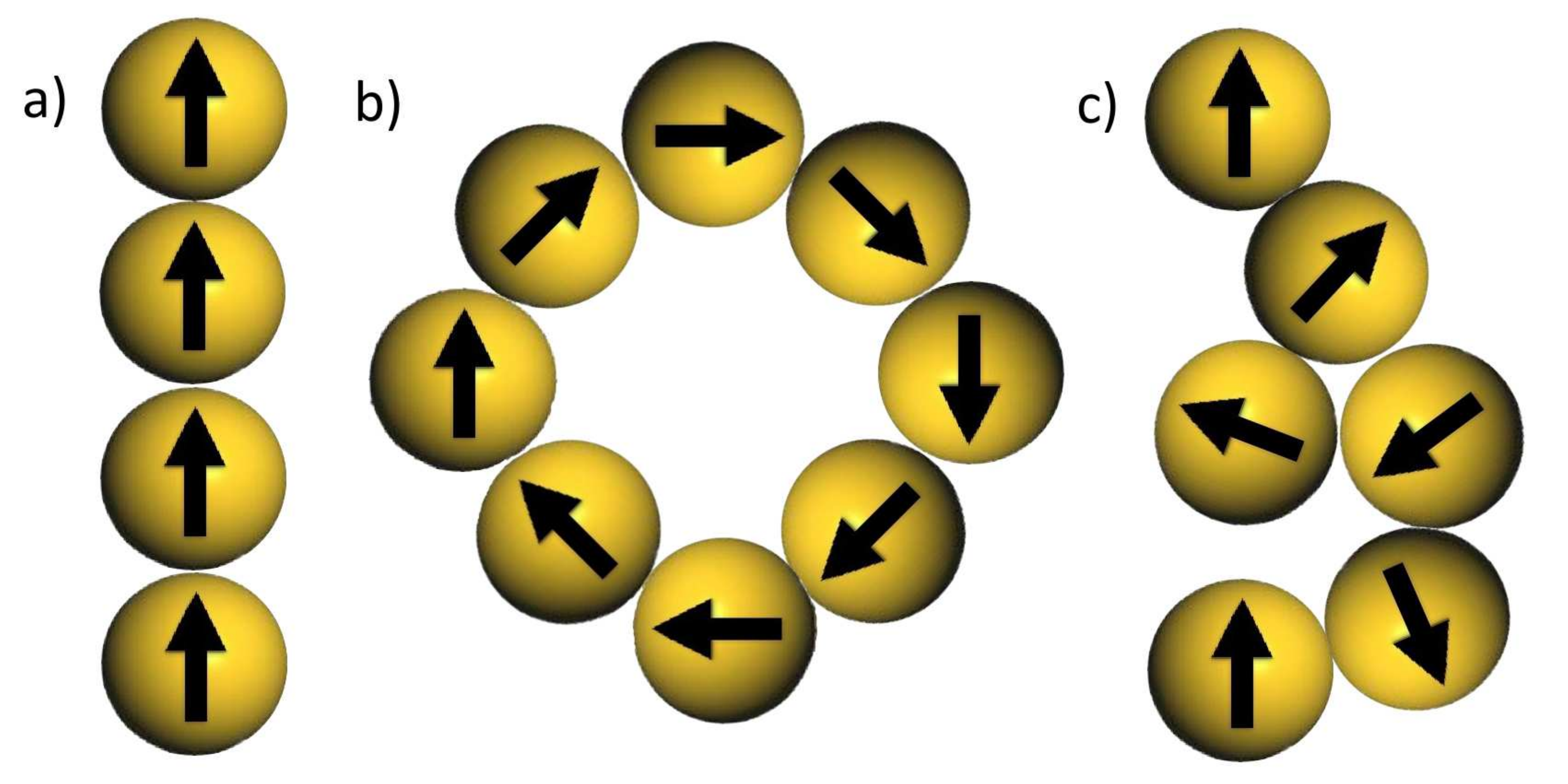}
\caption{The lowest energy state for dipolar particles in a) a moderate external field, b) in no field, and c) a messy, high-temperature agglomerate of dipolar particles.}
\label{fig:LoopChain}
\end{figure}

\begin{figure}[h]
\centering
\includegraphics[width=0.8\textwidth]{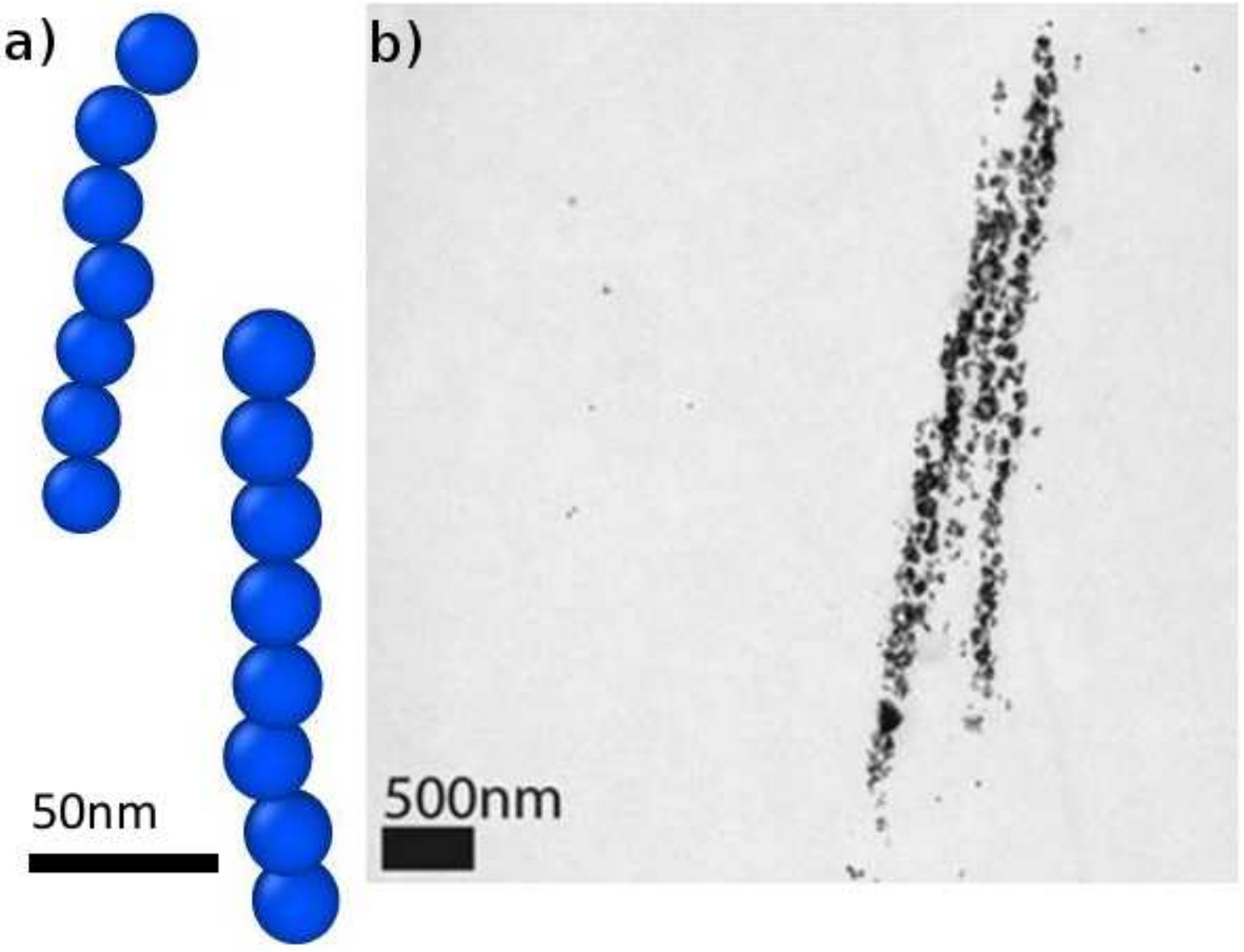}
\caption{Agglomeration of 22~nm diameter magnetite nanoparticles in water with 5~nm long PEG ligands, and with a 0.27~T applied field, predicted by theory a), and experimentally measured b).\cite{Saville:2013hf} Theoretical results predict single particle width agglomerates whereas experimental results show agglomerates with multiple particle widths and messiness even in the presence of an applied field.}
\label{fig:expVT}
\end{figure}

As an example of the difference between theory and experiment, in Fig.~\ref{fig:expVT} we show the predicted agglomeration (panel a) in comparison with an experimental result (panel b) in an applied field, taken from Ref.~\cite{Saville:2013hf}. The simulation parameters used are chosen to match the experiment and are given in the caption. In particular, these magnetite particles are large enough and the ligand lengths are short enough that magnetic interactions \emph{should} dominate thermal and electrostatic forces that would favor spherical clumps. Although the simulation is run with just a few particles to minimize computational time, the contrast in shapes is clear. In panel a), single-width particle chains that are very straight form while in experiment the chains are dendritic and are not straight. Also, some rounded clumps occur that can be seen by the dark, compact blobs that are actually comprised of many particles. (Note that the two chains in panel a) appear to be different sizes due to perspective and the one on the left is in the background.) The details of the simulations will be provided below but this figure is included here to motivate why current Langevin dynamics simulations need to be altered to produce realistic results.

We have performed Langevin dynamics simulations of a small number of single-domain ferromagnetic particles in a liquid using a code built in Fortran, plus using LAMMPS \cite{lammps}, to investigate this question. We characterize the formed agglomerate shapes using normalized length and the difference in angle between adjacent particles in order to compare the simulation results. Introducing thermal buffeting alone does not create the degree of messiness seen in experiment. Nor does a distribution of particle sizes. The electrostatic van der Waal's interaction between particles needs to be increased in strength well beyond accepted literature values to dominate the magnetic interactions and induce isotropic or spherical clumping, as will be discussed below. Instead, we find that if there is \emph{stickiness or ligand bridging} between particles then they may connect in ways that do not minimize the dipole-dipole interaction and that nucleates a clump that will not be a perfect chain, like that seen in Fig.~\ref{fig:expVT}(a). This stickiness must be strong enough so that the timescale for the particles to reorient to minimize their dipole-dipole interaction is longer than the time it takes for another particle to join the clump. We use this fact to estimate the size of a dimensionless, phenomenological stickiness-parameter $c$ that can theoretically take on values between 0 (no bridging between ligands) and 1 (irreversible bridging or sticking together on impact). We show how $c$ can be implemented in Langevin dynamics simulations to efficiently account for the effects of ligand bridging.

Our hypothesis is consistent with previous simulation results. When magnetic nanoparticles are treated as individual bodies without stickiness between them, then only chains that are one-particle-width form \cite{Wang02}. However, when coarse-graining techniques are used to speed-up simulations -- which effectively locks particles together as if bridging is instantaneous on impact and irreversible -- then much thicker and more disordered agglomerates are produced \cite{Satoh96}. Here the term coarse-graining refers to simulations in which particles that touch are assumed to be locked together forever and clumps move as one rigid object. Since they are stuck together and cannot slide over one another nor pull apart, this corresponds to an infinite amount of stickiness between touching particles. Furthermore, when particles are forced to remain neighboring (an analogue to introducing stickiness as we have done), then a variety of agglomerate shapes form depending on the competition between magnetic and thermal energies.\cite{pshen}

It should be noted that entropic forces also can lead to thicker chains of particles forming for large numbers of superparamagnetic \cite{Faraudo16} and ferromagnetic nanoparticles but are not large enough to produce the amount of disorder seen, for example, in Fig.~\ref{fig:expVT}(b). Moreover, Langevin dynamics simulations contain all the physics needed to recover results of statistical mechanics so this mechanism is built into our theory.

When magnetic nanoparticles form agglomerates, their function is very different compared to isolated particles. Therefore, this work is very important in order to create realistic clumps of particles to study. For example, particles in clumps may generate more or less heat for magnetic hyperthermia cancer treatment compared to isolated particles, depending on the cluster shape \cite{saville2014formation, Serantes2014assemblies, Carrey14} and this has lead to intense study over the last three years. Theoretical studies have used artificially-shaped clusters to demonstrate this point, such as rectangular prisms,\cite{saville2014formation, Serantes2014assemblies} perfect chains, spheres\cite{Carrey14} and fractal shaped clumps.\cite{Ondrej} Moreover, these studies have guessed or randomly distributed the easy anisotropy axes of particles within clusters. By simulating here the formation of clumps that actually match experiment, without the need to approximate easy axes directions, researchers will gain a better understanding of how agglomerate shapes and their internal magnetic structure affect their function. 
This work represents a first step in understanding the mechanisms that give rise to realistic shaped agglomerates in that only 100 particles are simulated to demonstrate that stickiness is important.

In section~\ref{MD} we will briefly describe the Langevin dynamics simulation involving single-domain ferromagnetic nanoparticles. We leave it to a later work to consider superparamagnetic particles where the magnetization within a particle may fluctuate as the agglomerate forms. In section~\ref{characterization} we describe the methods used to characterize the shapes of the formed agglomerates in order to compare with experiment. In section~\ref{estimates} we estimate the strength of the phenomenological stickiness parameter $c$ such that messy clumps form for various types of particles. Then in section~\ref{results} we discuss the simulation results. In section~\ref{concl} conclusions and future work are presented.



\section{Langevin dynamics simulation}
\label{MD}

The dynamics of each particle $i$ is governed by Newton's second law, expressed as the set of four three-dimensional, first-order, stochastic differential equations. Firstly, we will write these equations without the possibility of ligand bridging and then describe how they are altered when stickiness is included phenomenologically. The four equations per particle without ligand stickiness are:
\bea
\dot{\vec{x}} (t) &=& \vec{v}_{i} (t),
\label{Langevin1} \\
m_{i} \dot{\vec{v}}_{i}(t) &=& \vec{F}_{i} \left[ \vec{x}_{j}(t) \right] - m_{i} \gamma_{t i} \vec{v}_{i}(t) + \vec{R}_{i}(t) ~~\equiv \vec{f}_i,
\label{Langevin2} \\
\dot{\theta}_{i}\hat{u}_i (t) &=& \vec{\omega}_i (t),
\label{Angular1} \\
I_i \dot{\vec{\omega}}_{i} (t) &=& \vec{T}_i[\vec{x_j(t)}] -I_i\gamma_{r i} \vec{\omega}_i(t)+\vec{Q}_i (t)   ~~\equiv \vec{\tau}_i (t),
\label{Angular2} 
\eea
where  the dots over variables indicate time derivatives. Here, for the translational motion described by Eqs.~(\ref{Langevin1}) and (\ref{Langevin2}), $\vec{x}_{i}(t)$ is the position, $\vec{v}_{i}(t)$ is the velocity at time $t$, $m_{i}$ is the mass of particle $i$, $\vec{F}_{i}$ is the total deterministic force acting on the particle due to the other particles (all the contributions to which we will discuss below), $\gamma_{t i}$ is the translational viscous coefficient of the particle moving through the fluid, and $\vec{R}_{i}$ is the random, fluctuating force that describes the effect of finite temperature on the particle. The total force at a time $t$ is written as shorthand $\vec{f}_{i}(t)$, as this will be convenient in what follows. 

In the rotational equations of motion~(\ref{Angular1}) and (\ref{Angular2}), the angle of rotation for a particle is $\theta_i$ around the direction $\hat{u}_i$, the angular velocity is $\vec{\omega}_i$, the torque acting on the particle due to the others is $\vec{T}_i$, the moment of inertia is $I_i$, $\gamma_{r i}$ is the rotational viscous coefficient of the particle, and the random torque is $\vec{Q}_i$. The total of all torques on particle $i$ at time $t$ is written as $\vec{\tau}_{i}(t)$. We will consider identical particles so we can drop the subscript $i$ for the mass $m$, the moment of inertia $I$ and the viscous friction coefficients $\gamma_r$ and $\gamma_t$. Note that the ratio $\gamma_r / \gamma_t = 10/3$ for perfect spheres,\cite{Wang02} which can be derived from Eqs.(\ref{gammaT}) and (\ref{gammaR}) that appear later. Also note that some articles choose to group $m \gamma_{t} = \Gamma_t$ and $I \gamma_r = \Gamma_r$ together but we prefer the $\gamma$ notation as it allows one to more easily understand the relationship between the rotational and translational viscosity.

When the particles are well-separated in space, then one may ignore the inertial terms in Eqs~(\ref{Langevin2}) and (\ref{Angular2}). However, when the particles get closer together (within a few diameters) then these terms dominate over the thermal fluctuations. In particular, the separation at which the random forces/torques are on the same order of magnitude as the deterministic forces/torques is critically important for determining the end agglomerate shapes that form. This is our reason to use Langevin dynamics to model these systems. In the future, our codes may be accelerated by ignoring inertial forces when particles are well-separated from others.

\begin{figure}[h]
\centering
\includegraphics[width=0.8\textwidth]{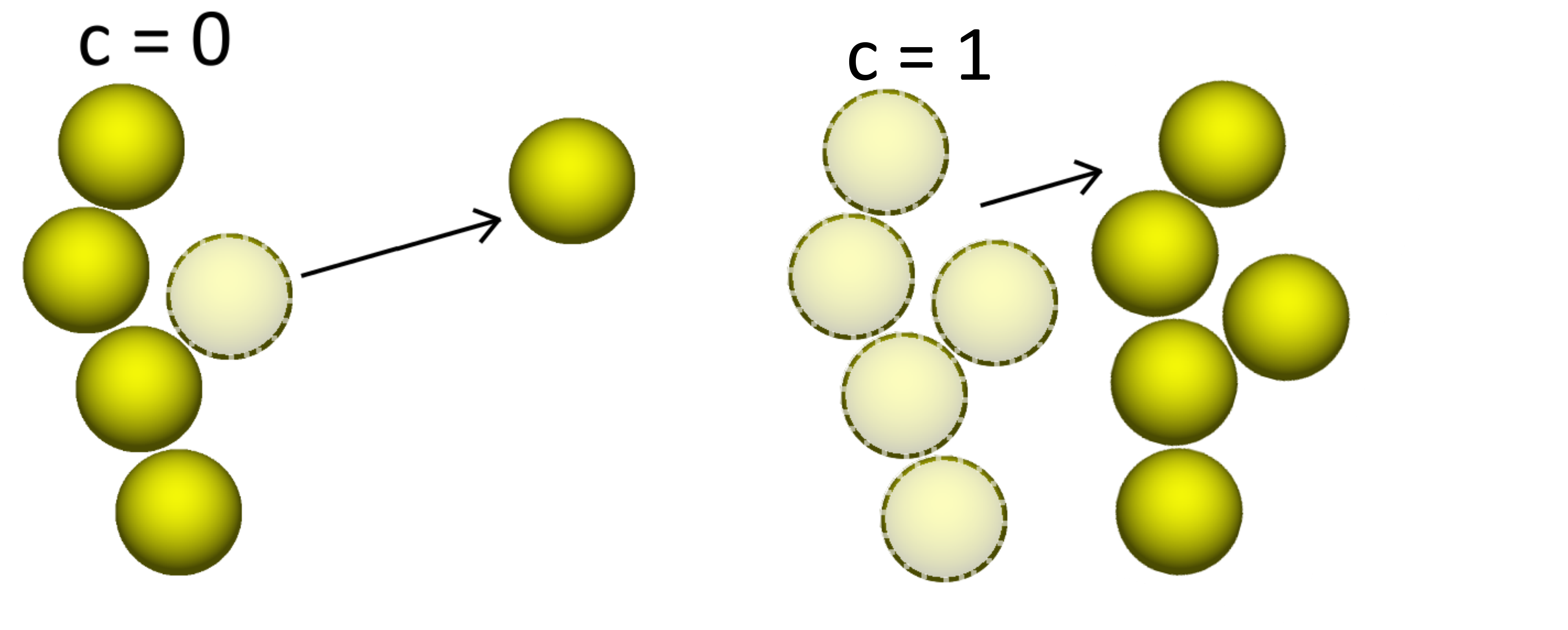}
\caption{A schematic showing the effect of no interparticle friction ($c = 0$) on the left, and infinite interparticle friction ($c = 1$) on the right.}
\label{fig:Cextremes}
\end{figure}

\begin{figure}[h]
\centering
\includegraphics[width=0.7\textwidth]{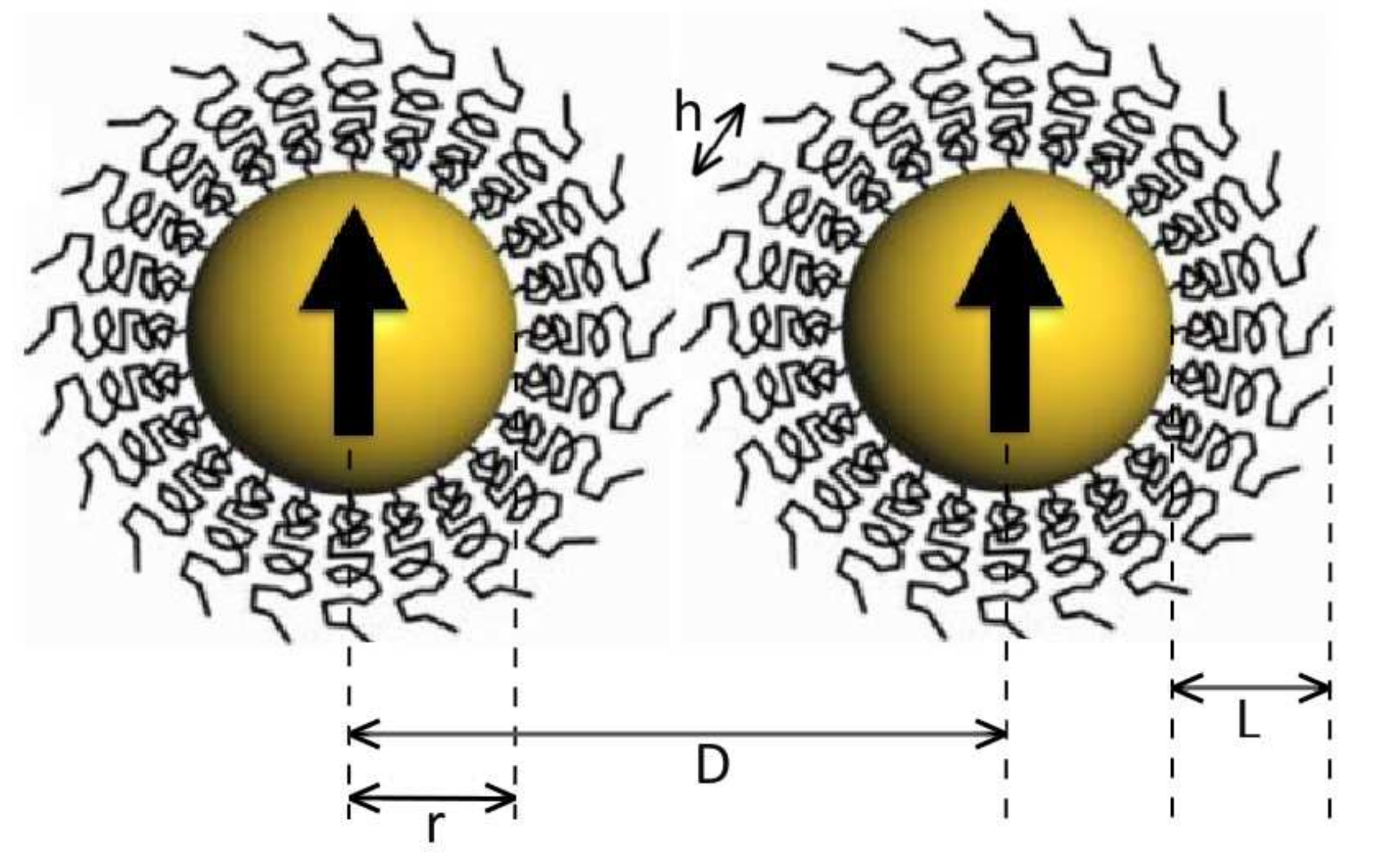}
\caption{A schematic of two dipolar particles where $\mathcal{D}$ is the interparticle distance, $r$ is the radius of the particle's magnetic core, $L$ is the ligand length and $h$ is the distance between ligand heads.}
\label{fig:LigandBrushes}
\end{figure}

Eqs.~(\ref{Langevin1})-(\ref{Angular2}) assume that particles may interact but due to thermal fluctuations always have a chance to act individually and escape the force-field of another, as shown schematically in Fig. \ref{fig:Cextremes} on the left. In contrast, assuming that once particles join together (their separation is less than some critical distance) they stay stuck (shown on the right in Fig. \ref{fig:Cextremes}), is unrealistic as it assumes that particles can never reorient. Instead, we use a model that allows situations between these two extremes. We define a dimensionless, phenomenological interparticle-stickiness parameter $c_{i}$, that is 1 if the particles are stuck together indefinitely once they touch and is 0 when particles always act as individual particles. We can then re-write Eqs.~(\ref{Langevin2}) and (\ref{Angular2}) as a weighted sum
\bea
m_{i} \dot{\vec{v}}_{i}(t) &=& \vec{f}_{i} (1-c) + \frac{c}{N_{agg}} \sum_{j=1}^{N_{agg}} \vec{f}_{j},
\label{Langevin2friction} \\
I_i \dot{\vec{\omega}}_{i} (t) &=& \vec{\tau}_{i} (1-c) + \frac{c}{N_{agg}} \sum_{j=1}^{N_{agg}} \vec{\tau}_{j},
\label{Angular2friction} 
\eea
where the sums are over the $N_{agg}$ particles that make up an agglomerate that particle $i$ is part of. One can see that if $c=1$, the agglomerate is treated as a single object, whereas if $c=0$ a particle is able to move independently, no matter how close it is to others and the agglomerate term vanishes. These two scenarios are drawn schematically in Fig.~\ref{fig:Cextremes}. These equations can be derived from Eqs. (\ref{Langevin2}) and (\ref{Angular2}) by finding the equations of motion for an agglomerate and solving for the weighting factor, requiring the system to remain conservative. This is shown in the Supplemental File for the translational equation. We will estimate the strength of $c$ in the next section. The term $c$ models the likelihood of the nanoparticles to stick due to contact of the ligand brushes, or ligand-bridging. Therefore, while the particles are within a critical distance of each other, namely the hydrodynamic diameter $\mathcal{D} \leq 2(r + L)$ (see Fig.~\ref{fig:LigandBrushes}), we apply this stickiness to all the forces acting on the particles. Here $r$ is the particles' magnetic core radius and $L$ is the mean ligand length.

Note that one could use a different value for the stickiness parameter $c$ in the translational and rotational equations of motion, Eqs.~(\ref{Langevin2friction}) and (\ref{Angular2friction}). However, the physical origin of the stickiness is ligand bridging and the fact that an elastic restoring force is experienced when ligands are stretched. The ligand or polymer strands are stretched in an almost equivalent manner when touching particles are translated or rotated and hence the restoring force in each case is on the same order of magnitude. For the phenomenological $c$ parameter, this means we can approximate it to be the same in both equations of motion. (In fact, we repeated simulations with a different value of $c$ in these equations and the results are unphysical. For example, particles are able to shear apart, but not translate apart.)

We will discuss the viscous and random forces/torques on a particle in more detail before describing the deterministic force contributions to Eq.~(\ref{Langevin2}). The drag coefficients in Eq.~(\ref{Langevin2}) and (\ref{Angular2}) are given by Stoke's formula \cite{Pathria,Landau}
\bea
m \gamma_t = 6 \pi \eta \mathcal{R},
\label{gammaT}\\
I \gamma_r = 8 \pi \eta \mathcal{R}^3,
\label{gammaR}
\eea
where $\eta$ is the viscosity of the fluid and $\mathcal{R}$ is the hydrodynamic radius of the particle. In turn, the viscosity of the fluid is related to the random force and torque of the fluid on a particle $\vec{R}_{i}(t)$ and $\vec{Q}_{i}(t)$. Writing just the statistical distribution governing the force relation, it is usually assumed to obey a Gaussian distribution centered about zero mean \cite{Kubo,Wang02}
\bea
\langle R_{\alpha}(t) \rangle &=& 0 
\label{meanR}\\
\langle R_{\alpha}(t) R_{\beta}(t+ t') \rangle &=& 6 m \gamma k_{B} T~ \delta_{\alpha \beta} ~\delta(t'),
\label{standdevR}
\eea
where $k_{B}$ is Boltzmann's constant, $T$ is temperature, $\alpha$ and $\beta$ represent the $x$, $y$ and $z$ components, $\delta(\tau)$ is the Dirac delta function and $\delta_{\alpha\beta}$ is the Kronecker delta function. Eq.~(\ref{standdevR}) indicates that the random force at time $t$ is completely uncorrelated with the random force at a time $t'$ later. This assumes that the timescales of fluid-particle interactions are much shorter than the characteristic timescales for the particle motion. Similar equations hold for the random torque $\vec{Q}_{i}(t)$.\cite{Wang02}

The deterministic forces represented by $\vec{F}_{i}$ in Eq.~(\ref{Langevin2}) include steric, van der Waal's and magnetic dipole-dipole and $\vec{T}_i$ are the associated torques. All of these forces depend on the configuration of the particles relative to each other. The van der Waal's attraction force between spherical particles is given by differentiating the energy of interaction \cite{hamaker}
\be
V_{Vpp} (\mathcal{D})  = - \frac{ A_{131} }{6} \left(  \frac{2 r^2}{\mathcal{D}^2-4 r^2} + \frac{2r^2}{\mathcal{D}^2 }  + \ln \left( \frac{\mathcal{D}^2 - 4 r^2}{\mathcal{D}^2} \right) \right).
\label{vdW}
\ee
Here $A_{131}$ is the Hamaker constant that depends on the material the particles are made from and their shape (in this case spherical), as well as the medium through which they interact. The radius of the particles is $r$ and the center-to-center distance between a pair of particles is $\mathcal{D}$.

The steric repulsive force prevents the ligand-coated particles from overlapping in space. It can be found by differentiating the steric potential between a pair of particles \cite{israel}
\be
V_{Spp} (\mathcal{D})  = \frac{100 r L^{3} }{(\mathcal{D}-2 r) \pi h^{3} } k_{B} T \exp \left( - \frac{\pi (\mathcal{D}-2r)}{L} \right),
\label{steric}
\ee
where $L$ is the average length of the ligands, $h$ is the mean distance between two ligand head groups on the surface of the particle, $k_{B}$ is Boltzmann's constant and $T$ is temperature. Some of these variables are shown in Fig.~\ref{fig:LigandBrushes}.

\begin{figure}[h]
\centering
\includegraphics[width=0.8\textwidth]{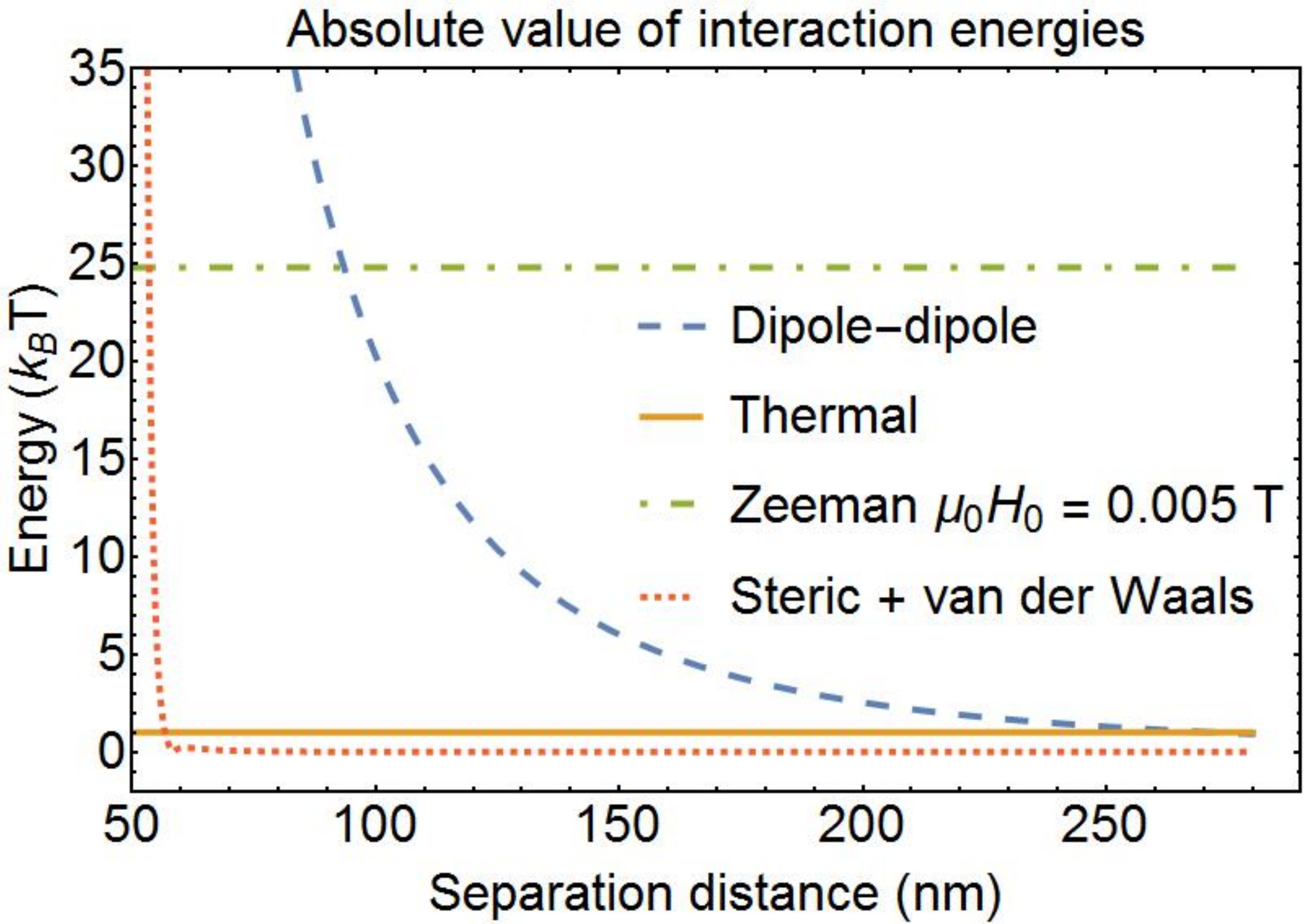}
\caption{A comparison of the absolute value of the relevant energies in the simulation. The blue (dashed) line is the dipole-dipole energy for tip-to-tail aligned dipoles as a function of distance, the orange (solid) line is the thermal energy, the green (dot-dashed) line is the Zeeman energy for a 5~mT applied field and the red (dotted) line is the sum of the steric and van der Waals energies as a function of distance for a 50~nm diameter magnetite particle with ligand lengths of 5~nm, a ligand headgroup spacing of 5~nm and a Hamaker constant of $2.9 \times 10^{-20}$ (dashed line).}%
\label{fig:LJPot}
\end{figure}

The interparticle interactions (van der Waal's and steric) can be added together to create a potential energy curve like the dotted curve shown in Fig.~\ref{fig:LJPot}. Here various energy contributions are plotted in units of $k_B T$ as a function of distance between cores. (We will discuss shortly the parameters used to generate this plot.) It can be seen that there is a minimum energy corresponding to an equilibrium center-to-center separation of roughly 62~nm. This energy is compared to the magnetic dipole-dipole energy for tip-to-tail aligned particles and the thermal energy, which has a value of 1 for all distances.



The final interparticle interaction to be considered is the dipole-dipole interaction. Each particle is assumed to have a magnetic dipole moment given by
\be
\vec{m}_{i} = \vec{M} \frac{4 \pi}{3} r^3,
\ee
where $\vec{M}$ is the saturation magnetization of the material and $r$ is the radius of the nanoparticle. It is assumed that a point dipole located at the center of the particle represents the magnetic properties and so two particles interact via an energy
\be
V_{ij}(\vec{r}_{i} - \vec{r}_{j}) = - \frac{ \mu_0 }{4 \pi |\vec{r}_{i} - \vec{r}_{j} |^3 }  \left( 3 (\vec{m}_{i} \cdot \hat{r}_{ij} ) (\vec{m}_{j} \cdot \hat{r}_{ij}) - \vec{m}_{i} \cdot \vec{m}_{j} \right),
\ee
where $\vec{r}_{i}$ is the position of particle $i$, the separation between particles is $\mathcal{D}= |\vec{r}_{i} - \vec{r}_{j}|$, and $\hat{r}_{ij}$ is a unit vector along the vector $\vec{r}_{i}-\vec{r}_{j}$.

There are additional hydrodynamic interactions that are neglected in this treatment. For example, there is a change in viscosity between approaching particles for small distances \cite{Bocquet}. However, due to the large magnitude of the magnetic forces as the distance between two particles decreases the effect of the change in viscosity is rather small. We also only consider particles with ligand coatings which adds a number of complications to the calculation of the distance-dependent viscosity for two nearby particles but generally makes this a weaker effect \cite{Leal}. Our stickiness parameter is independent of certain hydrodynamic/viscosity interactions.


There are many numerical routines for solving the equations of motion (\ref{Langevin1})-(\ref{Angular2}).\cite{AllenBook} We choose to use a modified leap-frog algorithm developed by Van Gunsteren and Berendsen \cite{VanGunst}. The random forces (torques) mimicking temperature are converted to random displacements and velocities (rotations and rotational velocities). 
The reader is referred to the original article for the equations as they are lengthy. The only change we make is the incorporation of the relevant forces for this study and the stickiness parameter $c$.

These algorithms were programmed into Fortran and were tested a number of ways. First, we turn off all forces apart from the viscous and random forces and verify that the equipartition theorem is satisfied. 
Secondly, we check that the particles acted on only by viscous and random forces should follow ``random walk" motion. Thirdly, we compared our results with the stickiness parameter $c=0$ to those calculated using LAMMPS and found agreement. 

We ran our fortran code on 25 Intel(R) Xeon 2.2 GHz processors over a duration of about 8 hours for 10 ms of simulation time with a timestep of 0.1 ns for 100 particles. While some previous works\cite{Faraudo16} have simulated a few thousand particles we chose to use only 100 and run multiple simulations to get statistical significance for the characterization parameters used. This is due to requiring a smaller timestep due to larger dipolar forces. The largest agglomerates we see are less than 50 particles and so we do not find the number of particles at this concentration to be a limiting factor. Our fortran code runs at a similar speed to the LAMMPS code, but has the advantage that our ligand-bridging effects characterized by $c$ are included. Both codes are run for 100 particles and with periodic boundary conditions so that results can be obtained in a reasonable amount of time. It is left to a future work to consider ways to speed up the calculation so that more particles and/or longer times can be simulated. We have also performed some simulations without periodic boundaries and found that there are not significant finite size effects in the simulation. 

We have chosen the following parameters for the simulation of magnetite particles with magnetization $M = 3.12 \times 10^{5}$~A/m, diameter $50$~nm, ligand length $L = 5$~nm, distance between ligand heads $h =  5$~nm and temperature $T = 298$~K. The surrounding fluid is hexane with a viscosity of $\eta = 2.97 \times 10^{-4}$~N s/m$^2$ and a corresponding Hamaker constant $A_{131} =29 \times 10^{-21}$~J. \cite{Faure2011} 
The timestep throughout the simulation is $\Delta t = 1 \times 10^{-10}$~s and the entire simulation duration is $10$~ms. Each simulation consists of 100 particles in a cube with side length $2,052$~nm using periodic boundary conditions, giving a volume concentration of magnetite of $7.58 \times 10^{-4}$. 

In addition, we have studied a second set of magnetite nanoparticles with 25~nm diameter and ligand lengths $L=10$~nm in water ($\eta=6.78 \times 10^{-4}$~Pa~s, $A_{131}=32.7 \times 10^{-21}$~J). These two different magnetite particles are chosen so as to study the contrast between large particles in a polar fluid where magnetic interactions should dominate, and particles near the superparamagnetic limit that are influenced more by thermal fluctuations. Aqueous environments are more important to study for biological applications but also have added complexity with double-layer electrostatic interactions. Magnetite is one of the few magnetic materials that are FDA approved for use in the body.

\section{Characterization of agglomerates}
\label{characterization}

Before discussing the simulations results, we describe the methods of numerically characterizing their results. To characterize the agglomerate shapes that form a number of parameters can be used. We have used a normalized length to simply describe the three dimensional shape of agglomerates. Additionally, to determine the amount of messiness in a chain we calculate an order parameter based on the difference in angle between adjacent particles, similar to Ref. \cite{Conde2008}. The method of calculation for these parameters are given below. 

To compare to experiment other characterization parameters may be of interest. For example the pair correlation function can be calculated \cite{pair} and compared to the structure factor which can be experimentally measured. Another quantity of interest is the fractal dimension which can be calculated from a two dimensional image obtained from theory or experiment.\cite{wozniak,liu,teichmann} Both the pair correlation and fractal dimension gain more value as the number of particles in the system becomes larger and so, we focus on the normalized length and angular order parameters. We did calculate the fractal dimension for all agglomerates following the methods of Refs.\cite{Ondrej,Etheridge}, however we do not present that data here due to a lack of consistency in the results. We attribute this lack of consistency to the fact that our simulations of 100 particles are just too small to get a value for the fractal dimension that can discriminate between messy and ordered clumps. Larger systems, plus more simulations (we do 8 for each parameter set) is needed.

A particle is identified as part of an agglomerate if it is separated by less than $2(r+L)$ from another particle (see Fig.~\ref{fig:LigandBrushes}). To calculate the normalized length of an agglomerate we identify the longest axis, defined by the maximum distance between two particles in the agglomerate 
\be
\mathcal{D}_{\textrm{max}}^{} = \max_{i,j} ~ \sqrt{(\vec{r}_i - \vec{r}_j)^2}  ,
\label{longAxis}
\ee
with $i$ and $j$ indexing all the particles in an agglomerate. The normalized length $L_n$ is then calculated from
\be
L_n = \frac{\mathcal{D}_{\max}^{}}{2(r+L)(N_{\textrm{agg}}-1)},
\label{normLen}
\ee
where $N_{\textrm{agg}}$ is the number of particles in the agglomerate. This length varies from 1, for a two particle agglomerate or a perfectly colinear chain where the particles have no overlap, to a minimum of approximately $0.14$. The minimum has been calculated for an arrangement of 20 particles placed in a hexagonal close-packed lattice, where the value $\mathcal{D}_{\textrm{max}}$ is kept to a minimum and particles cannot overlap. The simulation allows particles to overlap slightly, modeling a meshing of the ligand brushes, which will typically decrease the maximum and minimum values for $L_n$.

The normalized length is only calculated for the longest chain in the simulation. Initially, $L_n \approx 1$ for all simulations because the number of particles of the first chain is two. For some values of $c$, agglomerated particles can escape and rejoin the agglomerate or join a different agglomerate. Additionally, when multiple agglomerates form simultaneously the measurement for the normalized length may switch from one agglomerate to another throughout the simulation duration. This adds some ``noise" to calculations of $L_n$ and the number of particles agglomerated.

Another characterization parameter calculated is the angular order in alignment of all particles in the simulation. This parameter gives an indication of the misalignment of the dipole moments in the largest clumps even when the clump is nearly colinear. The angular order parameter is
\be
\Delta \theta = \frac{1-\frac{\vec{m}_1\cdot \vec{m}_2}{m^2}}{2 N_{\textrm{agg}}}
\label{qq}
\ee
where the values lie between 0, for clumps with colinear particles, and 1, for clumps with very disordered and misaligned particles.

The final characterization technique we used was to calculate the dipolar energy for an agglomerate. The calculation is 
\be
\Delta E =-\frac{1}{2 N_{agg}^2 E_{norm}} \sum_{i=1}^{N_{agg}} \sum_{j=1}^{N_{agg}} \frac{\mu_0}{4 \pi \mathcal{D}^3}(\frac{3 (\vec{m}_1 \cdot \vec{\mathcal{D}})( \vec{m}_2 \cdot \vec{\mathcal{D}})}{\mathcal{D}^2} - \vec{m}_1 \cdot \vec{m}_2)
\label{dE}
\ee
where
\be
E_{norm} =\frac{3 \mu_0 |m|^2}{4 \pi (2(r+L))^3} 
\label{En}
\ee
where the sums go over all 100 particles in the system excluding $i = j$. This calculation helps determine how messy all particles in the simulation are regardless of the clump size or linearity of chains. So, both a loop and a chain may have small energies when the particles are in a tip-to-tail orientation while a somewhat linear chain may have a larger energy if the particles don't line up tip-to-tail but are stuck in the initial contact configuration.

\section{Estimates for the stickiness parameter $c$}
\label{estimates}

In this work we explore the agglomeration behavior of magnetite nanoparticles with ligand coatings where the possibility of ligand bridging is modeled using the parameter $c$. Here, we estimate the value of $c$ for which the behavior in the simulations should transition from long chains to clumped agglomerates. We do this by comparing the time it takes a particle to diffuse ($t_{diff}$) to another, to the time it takes it to rotate so that the magnetic dipolar interactions are minimized ($t_{rot}$).

A characteristic rotation time for particles in a fluid experiencing a net torque is \cite{Berkov}
\bea
t_{{rot}} \sim \frac{8 \pi \eta R_{{hyd}}^3}{(1-c) \tau_{{total}}},
\label{rotationTime}
\eea
where the total torque $\tau_{{total}}$ a particle in an agglomerate experiences is decreased due to stickiness by the term $(1-c)$, consistent with Eq.~(\ref{Angular2}), and neglecting torques from far away particles (ie. those not in the agglomerate). $R_{hyd} \sim (r+L)$ is the hydrodynamic radius of the particle. This formula comes from the equation for the torque needed to keep a sphere rotating with uniform velocity in a still, infinite, viscous fluid.\cite{Lamb}

This can be compared to the expected time for agglomeration based on the density of particles and the thermal energy. If we assume that the total time $t_{total}$ for a particle to travel to another particle or an agglomerate is equal to the time to diffusively travel to the region close to another particle $t_{diff}$, plus the time to travel from the near-particle region (defined as where the magnetic interaction energy is equal to the thermal energy) to touching due to magnetic forces $t_{mag}$, then we get
\bea 
t_{{total}}^{} = t_{{diff}}^{} + t_{{mag}}^{}.
\label{diffusionTime}
\eea

The time for a particle to diffusively travel a distance $x$ is estimated using the fluctuation dissipation theorem result as \cite{Pathria}
\be
t_{diff} = \frac{x^2 6 \pi \eta R_{hyd}}{k_B T}.
\label{tdiff}
\ee
We estimate $x$ using the mean distance between the 100 particles initially in the simulation volume, minus the distance at which the magnetic interaction overwhelms thermal effects (the near-particle distance $r_{mag}\sim270$~nm for the 50~nm diameter particles). Therefore we have
\be
x = \left( \frac{V}{100} \right)^{1/3} - r_{mag}.
\label{xxx}
\ee
Substituting our parameters and Eq.~(\ref{xxx}) into Eq.~(\ref{tdiff}) finds a diffusion time on the order of 1~ms for the 50~nm particles.

In comparison, the time for a particle to travel from $r_{mag}$ to touching is \cite{Berkov}
\bea
t_{mag} = \int_{r_{mag}}^{r_{mag}/2} \frac{6 \pi \eta R_{hyd}}{F_{magnetic}(r) } dr,
\label{Eq:tmag}
\eea
where $F_{magnetic}$ is the magnetic force felt by the particles moving together. An estimate using our parameters is that $t_{mag}\sim 0.4$~ms. This means that $t_{total}$ is on the order of 1.4~ms. Note that this is an absolute upper estimate and agrees well with our total simulation time to see all particles agglomerate, which is 10~ms. The estimates do not account for the fact that the first particles have a probability to come into contact more quickly and as soon as one agglomerate nucleates, other particles are drawn towards it reducing the times. However, the estimates are useful to give analytic dependencies between variables.

When the time for a particle in an agglomerate to rotate so that its magnetic moment points in the local field direction, is larger than the average time for another particle to join the agglomerate, we expect a messy, noncolinear agglomerate to form. The new particle may stabilize the noncolinear shape. One may imagine an incoming particle to have been affected by thermal translations and rotations as well as experiencing the complicated superposition of dipolar fields from all particles on its approach, which is why its magnetic moment is not automatically aligned with the field of the agglomerate it joins. The condition for messy agglomerate shapes is therefore $t_{total} < t_{rotation}$.
Rearranging for $c$ and ignoring $t_{mag}$ to simplify the math, one obtains
\be
c > 1- \frac{4 k_{B} T }{3  \tau_{total} } \frac{R_{hyd}^{2} }{x^2}.
\label{cEstimate}
\ee
Then for a given particle concentration and temperature and particle size, we can estimate a stickiness parameter required to achieve noncolinear agglomerate shapes. Note that for larger thermal energies, less ``stickiness" (lower $c$ values) are required for messy clumps to form, as to be expected. Also note here that the fluid viscosity is absent as both $t_{rot}$ and $t_{diff}$ are linearly dependent on viscosity and the term cancels.

Using the parameters for the 50~nm diameter magnetite nanoparticles and assuming a magnetic torque $\tau_{total}$ for two particles in contact with an angle of $5^{\circ}$ between their dipole moments, Eq.~(\ref{cEstimate}) gives
\bea
c > 0.995
\eea
for messy agglomerates to form. This value can vary up to $c > 0.999$ by assuming the angle between dipoles is $90^{\circ}$ before they begin to rotate to minimize the dipole-dipole energy. Therefore, in our Langevin dynamics simulations we expect to find a normalized length of $L_n\approx 1$ and an angular order $\Delta \theta \approx 0$ for agglomerates when $c < 0.995$ and a decreasing normalized length for increasing values of $c$. 


In the case where an applied field is present we can account for the applied field in the estimate for $c$ by adding this to the torque experienced by a particle. In this case we get a new upper bound for the time required for a particle to rotate into the local field. This gives
\bea
c > 0.999
\eea
for messy agglomerates of 50~nm diameter magnetite particles with 5~nm ligand lengths to form in the presence of an applied field of $\mu_0 H_0 = 5$~mT.

We focus on this system because it is a good example of one for which most simulations predict linear chains of particles will form, but in experiment clumps are seen to form (see Fig.~2). For smaller particles, with long ligands, the magnetic interaction is effectively turned off and $c=0$ (ie. regular Langevin simulations) work well to predict the messy agglomerates that form. Our estimate for $c$ captures this behavior too, as is described below. 

We have also performed the above calculations for smaller magnetite particles with a diameter of 25~nm and 10~nm ligand lengths with the same volume concentration at a temperature of 310~K in water. In this case we get an estimate of 
\bea
c > 0.8
\label{csmall}
\eea 
with no applied field present and 
\bea
c > 0.99
\label{csmallF}
\eea
with an applied field of $\mu_0 H_0 = 5$~mT.

\begin{figure}[h]
\centering
\includegraphics[width=0.5\textwidth]{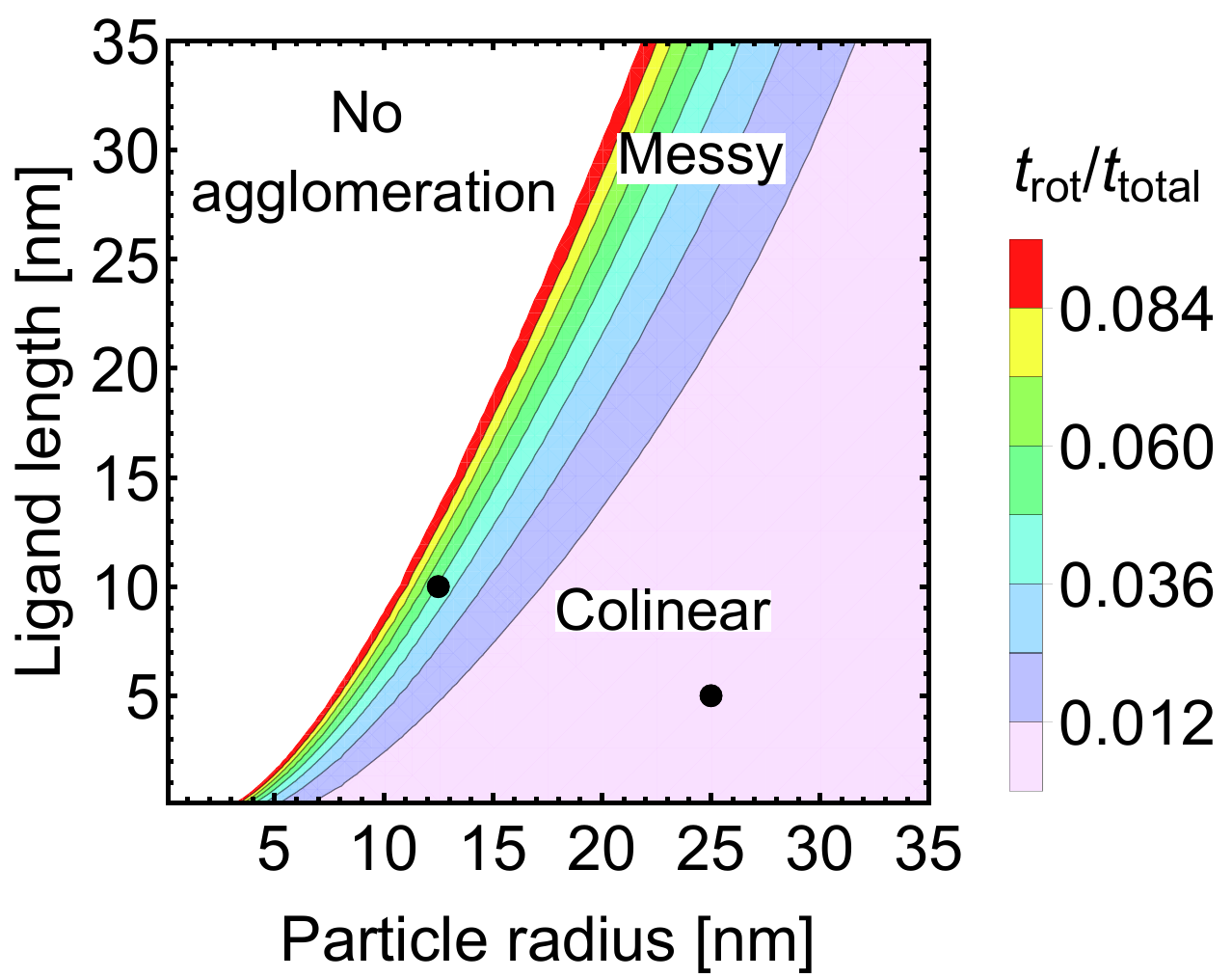}
\caption{The contour plot shows the ratio of $t_{rot}$ to $t_{total}$ with no stickiness coefficient ($c = 0$). The legend on the right-hand-side shows what values the colors represent. The white area on the left indicates the region where the thermal energy is larger than the sum of dipolar, steric, and van der Waals energies so agglomerates do not form. The red region indicates regions where agglomerates tend to be messier while the pale area on the right indicate agglomerates that are colinear. The two black dots are the parameters for the results shown in this manuscript.}
\label{fig:pD}
\end{figure}

The method used to analytically estimate $c$ can be extended to look at the behavior of agglomerating systems over a wide range of parameters, whereas we have run simulations of just two types of particles here. In Fig.~\ref{fig:pD} we compare the ratio of $t_{rot}/t_{total}$ (scale bar on right) as a function of particle radius $r$ and ligand length $L$ with no stickiness ($c = 0$), using a contour plot. This is done using Eq.~(\ref{rotationTime}) and (\ref{diffusionTime}). Particles in the red areas of Fig.~\ref{fig:pD} will form messier agglomerates whereas particles in the pale areas (bottom right) will form linear agglomerates. These results can be easily understood by considering the magnitude of the torque induced by the agglomerate on the considered nanoparticle. When the ligand coating is large (top), the separation between particle centers is larger, decreasing the magnetic torque. Similarly, when the magnetic dipole moments of the particles are small (left) the magnetic torque is again small. In these cases the particles reorient into a tip-to-tail configuration more slowly and noncolinear agglomerates are more likely to be formed. The right side of the graph shows larger radius nanoparticles which approach the multi-domain limit, above which the calculation is no longer valid. The assumptions also break down for particles on the far left side of the graph that are very small and display superparamagnetic behavior. Note that the two particle sizes we have chosen to study (25 and 12.5 radius particles) deliberately lie in two very different regions of this phase diagram, as shown by the two dots drawn on Fig.~\ref{fig:pD}, and so very different behaviors are expected.

When the particles are sufficiently small or ligand coatings are sufficiently large and there is no applied magnetic field, the thermal energies may be larger than the attractive magnetic and van der Waals energies. In this case we do not expect (stable) agglomerates to form and our estimate above is no longer valid as it assumes that particles experience experience deterministic torques. The region where agglomeration is not expected, regardless of $c$ value, is shown in white in Fig.~\ref{fig:pD}.

\section{Simulation results}
\label{results}

\begin{figure}[h]
\centering
\includegraphics[width=.7\textwidth]{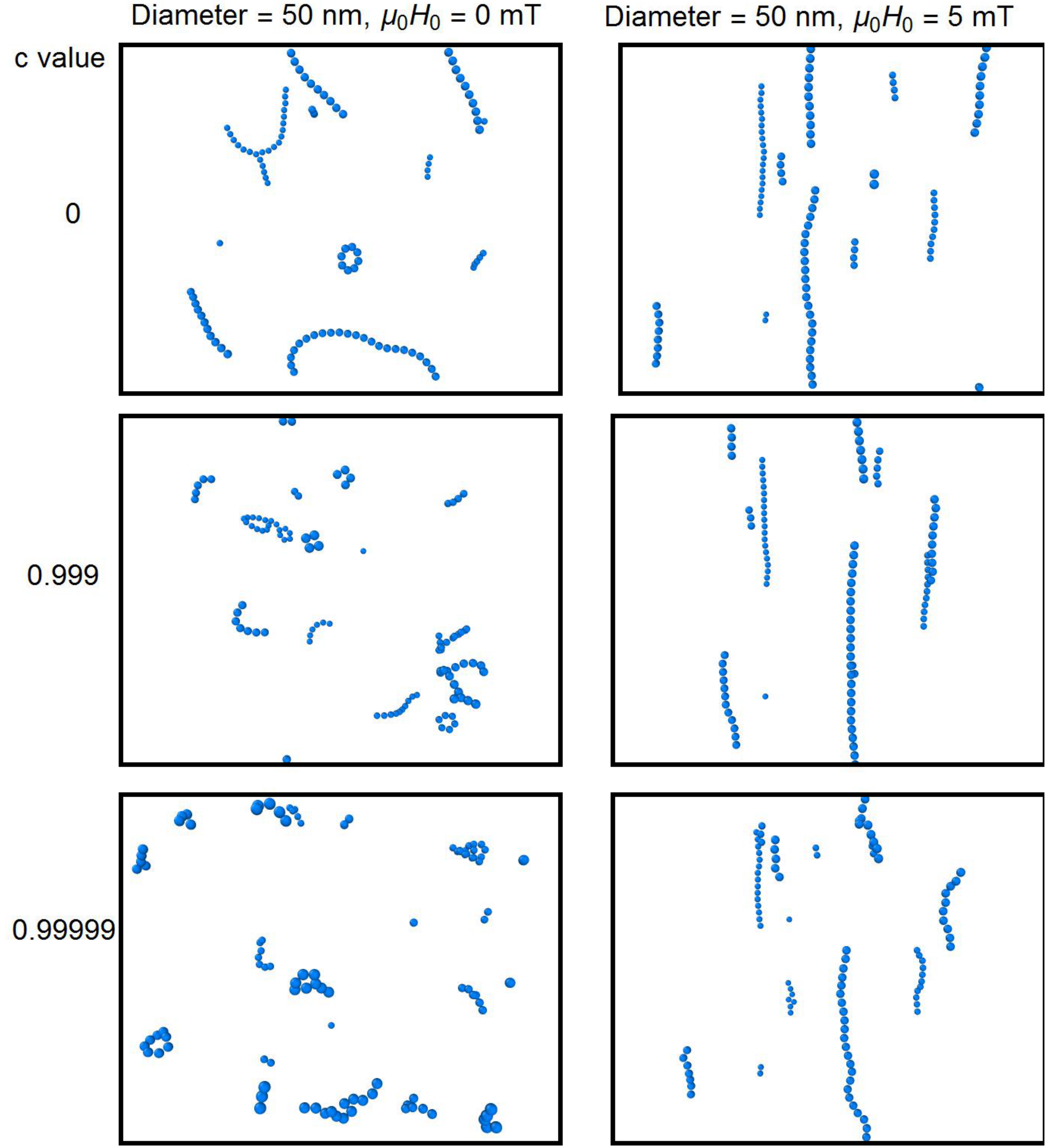}
\caption{A table of representative agglomerates formed in 0 field (left) and in an applied field of $\mu_0 H_0 = 5$~mT (right) for multiple values of $c$, for 50~nm diameter magnetite nanoparticles with 5~nm ligand coatings in hexane.}
\label{fig:shapes}
\end{figure}

In Fig.~\ref{fig:shapes} we show some of the simulation results for the 50~nm diameter magnetite nanoparticles with 5~nm ligands in hexane. Panels on the left show results in zero field after 10~ms and panels on the right show the results for an applied field of $\mu_0 H = 5$~mT ($\mu_0 m H_0=25 k_B T$ so this is a strong field). Three different values of the stickiness parameter $c$ are illustrated. When the particles cannot stick together due to ligand bridging ($c=0$, top panels) the particles form single-particle width chains due to dipolar interactions. In the top, left panel a loop can be seen that minimizes the magnetic dipole-dipole interaction, without an applied field.

As the stickiness parameter $c$ is increased to the critical value of 0.995 that we predicted earlier (middle and bottom panels) then the single-width chains are seen to transition to messier, clumpier agglomerates when there is no applied field (left panels). When there is an applied field in the vertical direction, then the chains remain due to the Zeeman interaction working with the dipolar interactions to favor chain formation, but the chains are seen to be less straight. The mean agglomerate size after 10 ms (averaging over 8 simulations of 100 particles) is roughly 10 particles, with slightly less particles per agglomerate at high values of $c$ due to agglomerates having weaker stray fields when they are disordered. We next use the characterization measures developed in Section~3 to quantify the changes in shape as a function of $c$.

\begin{figure}[h]
\centering
\includegraphics[width=1\linewidth]{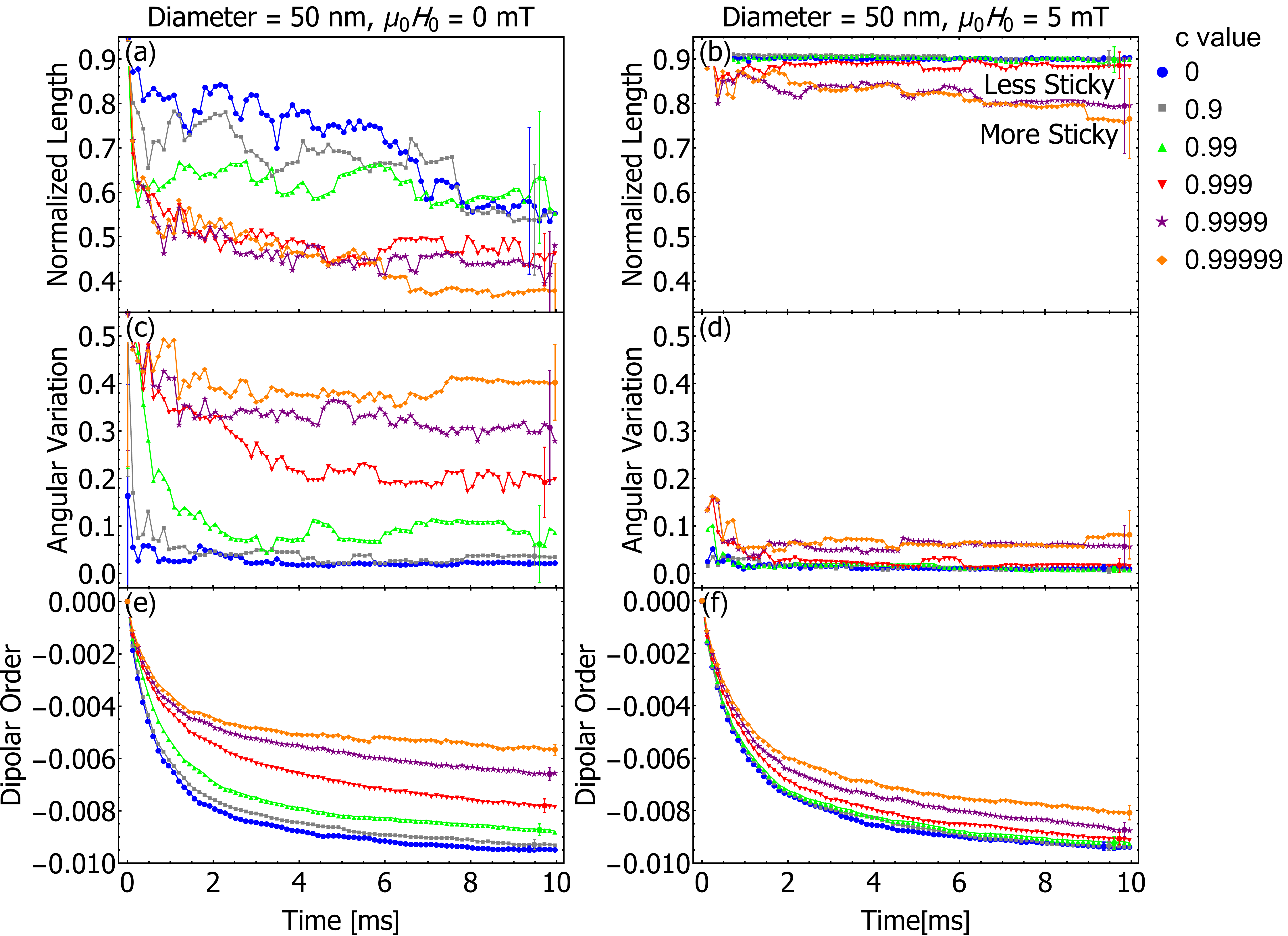}
\caption{In (a) the normalized length $L_n$ versus time with no applied field and in (b) the same for an applied field of $\mu_0 H_0 = 5$~mT . In (c) the angular variation versus time for multiple values of $c$ in no applied field and in (d) the same for an applied field of $\mu_0 H_0 = 5$~mT . In (e) the dipolar order versus time for multiple values of $c$ in no applied field and in (f) the same for an applied field of $\mu_0 H_0 = 5$~mT. In all plots a standard deviation is shown for a few representative values as an error bar.}
\label{fig:Ln}
\end{figure}

Fig.~\ref{fig:Ln} shows the calculated characterization parameters of agglomerates as a function of time for multiple values of $c$, in the absence of an external field on the left and in a field on the right. Different values of the stickiness coefficient $c$ are labeled by different colored symbols and a legend is given to the right of the panels. Each data point is calculated using an average from eight simulations and standard deviations are shown by the vertical lines at the right edge of each plot. In panels (a) and (b) the normalized length $L_n$ is plotted. In panels (c) and (d), the angular variation is plotted. In panels (e) and (f), the dipolar order parameter is plotted as a function of time.

The first thing to notice in Figure~\ref{fig:Ln} is that all of the characterization parameters have generally settled into some steady-state within the 10~ms time of the simulation. This (along with long computational times) is our reason for simulating this period of real time. Next, one can look at the values for normalized length (panels (a) and (b)) and see that $L_n$ decreases with larger values of stickiness $c$. This is as we expected because stickier particles are expected to form more compact and less linear agglomerates. Note that the initial value for $L_{n}$ at time $t=0$ is nearly 1 as initially all particles are separated in the simulation. The normalized length is longer for agglomerates formed in a field (compare panels (a) and (b)), and this matches the observations of Fig.~\ref{fig:shapes}. 

Note that the normalized length is not close to 1 in no field and when $c=0$ and the standard deviations are large because the average over all agglomerates simulated includes straight chains and closed loops, as shown in Fig.~\ref{fig:shapes}. Therefore, in zero field this characterization method is not as good as the others in determining the agglomerate regime we are in. It does, however, have the advantage of being simple to compute from experimental images of agglomerates and therefore is a good metric for comparing theory to experiment.  

The formation of messy agglomerates in an applied field with an energy of $25$ times $k_B T$ is a somewhat surprising result and so we elaborate on the process of agglomerate formation in a field. When $c$ is large, as the particles approach each other the dipole-dipole interaction becomes stronger than the Zeeman interaction at a distance of $ \mathcal{D} \approx 93$~nm. Near this distance a particle begins to rotate into the dipole field from the approaching particle. As agglomeration occurs, typically, the particles have not fully rotated into a tip-to-tail orientation, but have a finite angle between both their respective dipole moments and the applied magnetic field. This creates a nucleation site for further noncolinear agglomeration as other particles and agglomerates approach. However, when $c$ is small the same process occurs but after agglomeration the dipole moments continue to rotate into tip-to-tail alignment as well as alignment with the applied magnetic field.

The angular variation (Fig.~\ref{fig:Ln}(c) and (d)) increases as the stickiness $c$ is increased. This again reflects that the ligand bridging between particles prevents the dipolar interactions from being minimized. When a magnetic field is applied (panel (d)) then the angular variation is dramatically reduced because most of the magnetic dipoles align with that field. For both of these parameters (normalized length and angular variation) the value $c = 0.999$ seems to represent a phase change from colinear to noncolinear agglomerates as predicted by the analytic estimates above in Section~4. Colinear structures have low angular variation and longer normalized lengths. Noncolinear or compact agglomerates have larger angular variations between dipole moments and a shorter normalized length. 

Finally, the dipolar order parameter -- plotted as a function of time in Fig.~\ref{fig:Ln}(d) and (f) -- is the smoothest parameter and is monotonically decreasing, indicating that the magnetic energy is always decreasing in the system. Initially the energy is close to zero as particles are randomly oriented. The energy is smaller for no ligand bridging ($c=0$, blue data) and is considerably larger for high stickiness ($c=0.999$ and larger, red data) reflecting that particles get stuck in higher energy states. Although this parameter has the smallest standard deviation and is the best indicator of the agglomerate regime (sticky versus free particles) that a system is in, it is very difficult to experimentally image the magnetization directions of nanoparticles in a fluid in order to compare simulations to real systems. One exception is using off-axis electron holography, which has been used to image Co nanoparticles forming chains or rings.\cite{Wei2011}

It should be noted that simulations were performed with the magnetic dipole moment set to zero in order to determine the role of the van der Waals interaction in the clumping process. As this is an isotropic interaction, it could account for noncolinear particle agglomeration. However, in order to see any stable clumps form, the Hamaker constant had to be increased by greater than a factor of 10 from the value we used, which takes it far beyond the range of possible values quoted in the literature.\cite{Faure2011} This suggests that while the van der Waals interaction does stabilize clumps in some instances, here it is not strong enough to overcome thermal forces.

\begin{figure}[h]
\centering
\includegraphics[width=.7\textwidth]{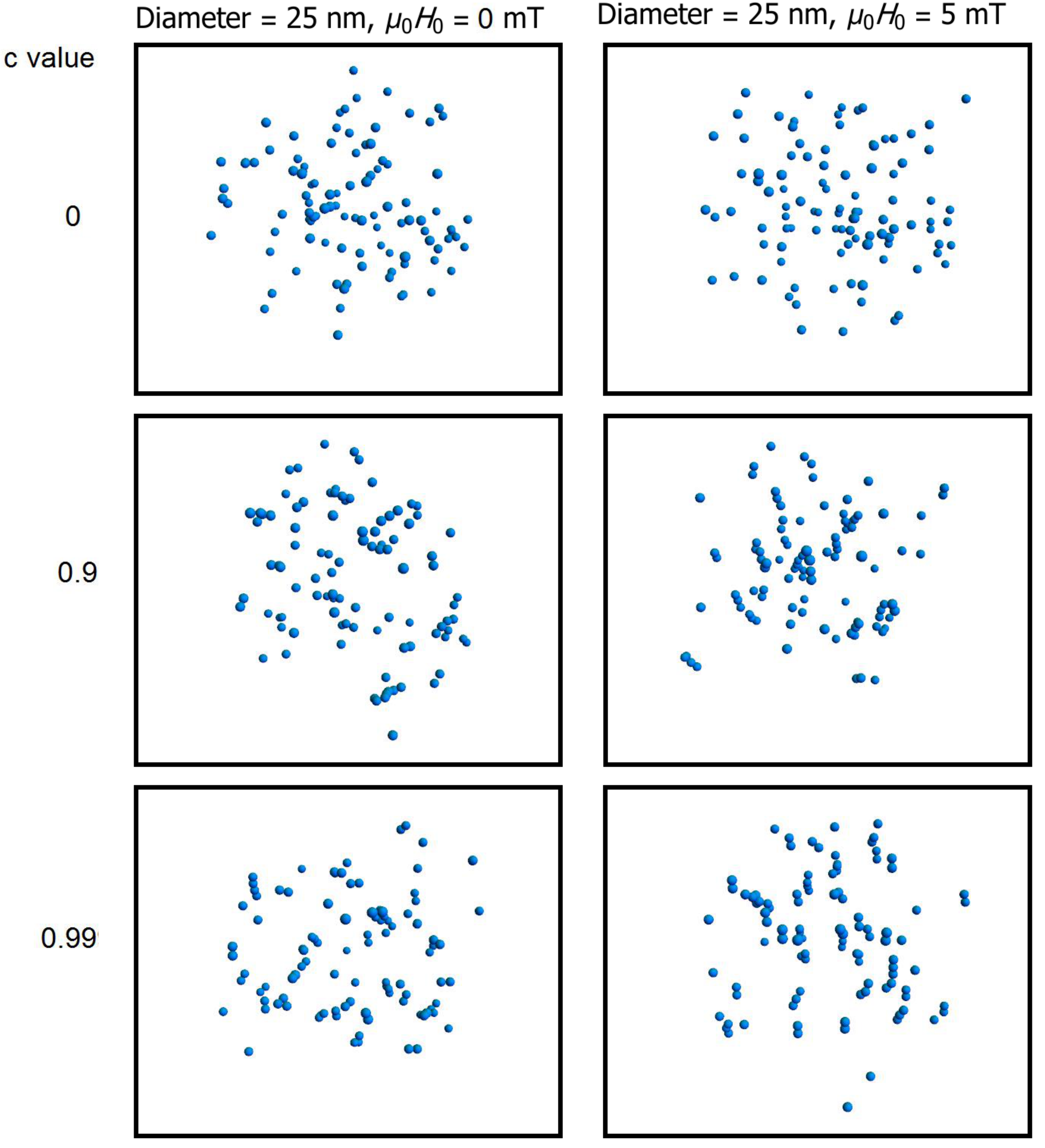}
\caption{A table of representative particle formations in 0 field (left) and in an applied field of $\mu_0 H_0 = 5$~mT applied upwards (right) for multiple values of $c$, for 25~nm diameter magnetite nanoparticles with 10~nm ligand coatings in hexane. Here the full hydrodynamic radius of the particles are shown to better see the formed aggomerates.}
\label{fig:smallshapes}
\end{figure}

The simulation results for the smaller 25~nm diameter magnetite nanoparticles with 10~nm ligands in water are shown in Fig.~\ref{fig:smallshapes}. Panels on the left show results in zero field after 10~ms and panels on the right show the results for an applied field of $\mu_0 H = 5$~mT. For the smaller particles we found very little clumping for the $c=0$ case as the thermal energies are large enough to move particles apart after they have come into contact for short periods of time. We therefore show representative particle positions after 10~ms for  values $c = 0.9, 0.99,$ and $0.999$ where some agglomeration was seen. For these cases the average agglomerate size is roughly 2 meaning that many particles are still unagglomerated, due to the thermal forces just described. For these smaller particles in no field we characterize the clumps with size greater than 2 particles as noncolinear for all these values of $c$, in agreement with the estimate in Eq.~(\ref{csmall}). This definition is based on the characterization metrics described below. 

\begin{figure}[h]
\centering
\includegraphics[width=0.9\linewidth]{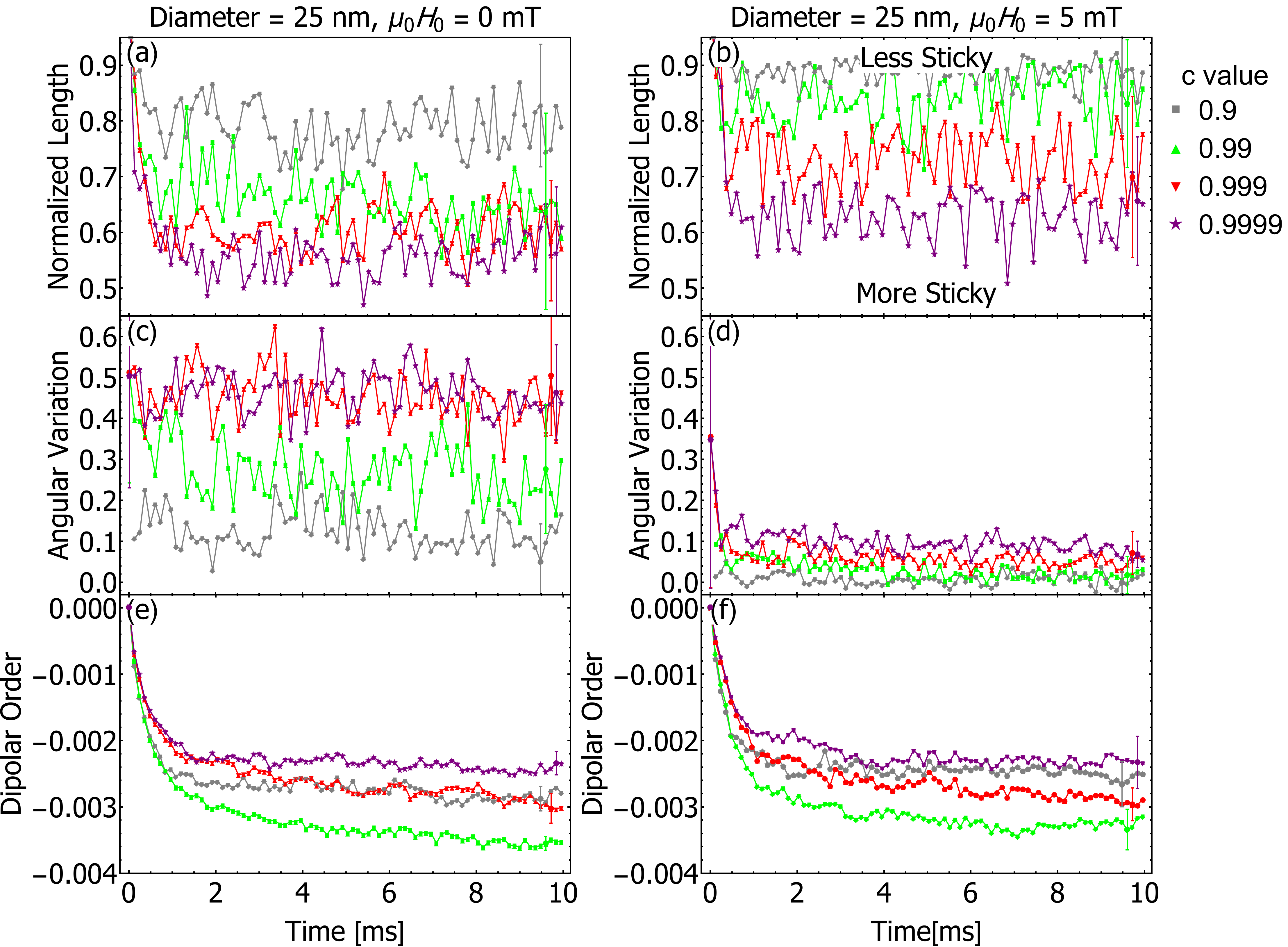}
\caption{In (a) the normalized length $L_n$ versus time with no applied field and in (b) the same for an applied field of $\mu_0 H_0 = 5$~mT . In (c) the angular variation versus time for multiple values of $c$ in no applied field and in (d) the same for an applied field of $\mu_0 H_0 = 5$~mT . In (e) the dipolar order versus time for multiple values of $c$ in no applied field and in (f) the same for an applied field of $\mu_0 H_0 = 5$~mT. In all plots a standard deviation is shown for a few representative values as an error bar.}
\label{fig:Ln2}
\end{figure}

Fig.~\ref{fig:Ln2} shows the calculated characterization parameters as a function of time for the 25~nm diameter particles, for multiple values of $c$. Results in the absence of an external field are on the left and in a field are on the right. Each data point is calculated using an average from 8 simulations and standard deviations are shown by the vertical lines at the right of each plot. Panels (a) and (b) show the normalized length, panels (c) and (d) show the angular variation and panels (d) and (f) show the dipolar order parameter. Note that the results are much less smooth than those for the 50~nm diameter particles because the average agglomerate size is close to 2 and so there are large fluctuations in all values.

We see in Fig.~\ref{fig:Ln2} that $L_n$ varies as $c$ is changed as expected, where smaller values of $c$ produce longer chains ($L_n$ closer to 1). Similarly, the angular variation increases as the stickiness, $c$, is increased. Finally, the dipolar order has curious behavior in that it at first decreases for increasing stickiness parameters (compare $c=0.9$ in gray and $c=0.99$ in green online) and then increases again (for example, see $c=0.9999$ in purple). This is because for small values of $c$ the ligand bridging can overcome thermal fluctuations and keep particles locked together for longer in configurations that lower the dipolar energy. But then for larger values of $c$, the particles can become locked together indefinitely in configurations that have higher dipolar energy. All these results show that for such small particles the effect of ligand bridging is small as thermal agitations play such a dominant role in the dynamics.

\section{Conclusion}
\label{concl}

The torques and forces on touching magnetic nanoparticles due to ligand bridging can be modeled using a phenomenological stickiness or frictional parameter $c$. Langevin dynamics calculations with $c$ included indicate a clear transition from linear to clumpy agglomerates forming as a function of $c$, for particles that have dipolar interactions that dominate over thermal forces. 

We estimate the critical value of $c$ using a ratio of the diffusion time of a free particle to the rotation time of a magnetic particle in an agglomerate. We find that for values of the stickiness parameter greater than $c = 0.99$, for 50~nm diameter magnetite particles with 5~nm ligand shells, the agglomerates are ``messy'' but linear chains or loops are formed when $c$ is smaller. This is in complete agreement with our simulation results.


Our calculation should be scaled up to larger volumes, more particles, and longer times to model real systems. But the fact that we can explain the occurrence of messy agglomerates is of great promise for calculating the properties of real magnetic nanoparticle systems. For example, it has been shown that the shape of agglomerates can dramatically alter the heat produced by magnetic nanoparticles for hyperthermia treatment \cite{saville2014formation, Serantes2014assemblies, Carrey14} but the agglomerates so far theoretically studied were all regular shapes (chains, rectangular prisms or spheres). Our simulation results allow for predictions to be made for much more realistic shapes that form in liquids.

Additionally, our calculations for the value of the stickiness coefficient should be compared to physical measurements of the ligand interaction strengths. While specific numbers are lacking for magnetite nanoparticles with PEG ligands, a number of experiments have been done for other surface/ligand combinations including PEG ligands. \cite{Sun2016,Xue2014,Dong2011} Among these we find binding forces ranging from 100 -- 700 pN over distances from 0 to 7~nm. This gives energies around $10^{-19} - 10^{-22}$~J which is comparable with the magnetic energies and thermal energies in this analysis. Therefore, the critical value for the stickiness coefficient $c$ that we found seems plausible but needs better experimental validation.

Future work may also add electrostatic interactions to the calculation. This can be done with the same method of incorporating stickiness between touching particles. A screened Coulomb potential can be used and appears to be important for ionic solutions containing magnetite nanoparticles. Recent work has shown that assembly happens at vastly different rates and with different shapes with slight changes in ionic strength.\cite{meff}

 \section*{Acknowledgement}
Support is acknowledged from the UCCS Center of the Biofrontiers Institute. D.~S. acknowledges financial support from Xunta de Galicia (Spain) under Plan I2C.



\bibliographystyle{model1a-num-names}
\bibliography{<your-bib-database>}

\begin{thebibliography}{00}


\bibitem{DobsonReview}
J.~Dobson.
\newblock Magnetic nanoparticles for drug delivery.
\newblock \emph{Drug Develop. Res.}, 67(1):55--60, 2006. DOI: 10.1002/ddr.20067


\bibitem{Fortin}
J.~P.~Fortin, C.~Wilhelm, J.~Servais, C.~M{\'e}nager, J.~C.~Bacri, and F.~Gazeau.
\newblock {Size-sorted anionic iron oxide nanomagnets as colloidal mediators
  for magnetic hyperthermia}.
\newblock \emph{J. Am. Chem. Soc.}, 129(9):2628--2635,
  2007. DOI: 10.1021/ja067457e
  
  \bibitem{WangLC}
  M.~Wang, L.~He, S.~Zorba and Y.~Yin.
  \newblock {Magnetically actuated liquid crystals}.
  \newblock \emph{Nano Lett.}, 14(7):3966-3971, 2014. DOI: 10.1021/nl501302s
  
  
  \bibitem{LeeLab}
  J.-J.~Lee, K.-J.~Jeong, M.~Hashimoto, A.~H.~Kwon, A. Rwei, S.~A. Shankarappa, J.~H. Tsui and D.~S. Kohane.
  \newblock {Synthetic ligand-coated magnetic nanoparticles for microfluidic bacterial separation from blood}.
  \newblock \emph{Nano Lett.}, 14(1):1-5, 2014. DOI: 10.1021/nl3047305
  
  \bibitem{Bonnemain}
  B.~Bonnemain.
  \newblock {Superparamgnetic agents in magnetic resonance imaging: physicochemical characteristics and clinical applications review}
  \newblock \emph{J. Drug Target.}, 6(3):167-174, 1998. DOI: 10.3109/10611869808997890
  
  \bibitem{Wei2011}
  A.~Wei, T.~Kasama and R.~E.~Dunin-Borkowski.
  \newblock {Self-assembly and flux closure studies of magnetic nanoparticle rings}.
  \newblock \emph{J. Mater. Chem.} 21:16686-16693, 2011. DOI: 10.1039/C1JM11916H
  
  \bibitem{Jund95}
P.~Jund, S.~G.~Kim, D.~Tom\'anek and J.~Hetherington.
\newblock Stability and Fragmentation of Complex Structures in Ferrofluids.
\newblock \emph{Phys. Rev. Lett.}, 74, 3049, 1995. DOI: https://doi.org/10.1103/PhysRevLett.74.3049
  
    \bibitem{Cheng} G.~Cheng, D.~Romero, G.~T.~Fraser and A.~R.~H.~Walker.
  \newblock Magnetic-field-induced assemblies of cobalt nanoparticles.
  \newblock \emph{Langmuir}, 21(26):12055-12059, 2005. DOI: 10.1021/la0506473
  
  \bibitem{rovigatti}
  L.~Rovigatti, S.~Kantorovich, A.~O. Ivanov, J.~M.~Tavares, and F.~Sciortino. 
  \newblock Branching points in the low-temperature dipolar hard sphere fluid. 
  \newblock \emph{J. Chem. Phys.}, 139(13):134901, 2013. DOI: http://dx.doi.org/10.1063/1.4821935
  
  \bibitem{Wang02}
Z.~Wang, C.~Holm and H.~W. M\"uller.
\newblock {Molecular dynamics study on the equilibrium magnetization properties and structure of ferrofluids}.
\newblock \emph{Phys. Rev. E}, 66:021405, 2002. DOI: https://doi.org/10.1103/PhysRevE.66.021405

  
  \bibitem{Saville:2013hf}
S.~L. Saville, R.~C. Woodward, M.~J. House, A.~ Tokarev, J.~
  Hammers, B.~ Qi, J.~ Shaw, M.~ Saunders, R.~R. Varsani, T.~G.
  St~Pierre, and O.~T.~ Mefford.
\newblock {The effect of magnetically induced linear aggregates on proton
  transverse relaxation rates of aqueous suspensions of polymer coated magnetic
  nanoparticles}.
\newblock \emph{Nanoscale}, 5:2152--2163, 2013. DOI:  10.1039/c3nr32979h

\bibitem{martinez}C.~ Martinez-Boubeta, K.~ Simeonidis, D.~ Serantes, I.~ Conde-Lebor\'an, I.~ Kazakis, G.~ Stefanou, L.~ Pe\~na, R.~ Galceran, L.~ Balcells, C.~ Monty, D.~ Baldomir, M.~ Mitrakas, and M.~ Angelakeris. 
\newblock Adjustable Hyperthermia Response of Self- Assembled Ferromagnetic Fe-MgO Core-Shell Nanoparticles by Tuning Dipole-Dipole Interactions. 
\newblock \emph{Adv. Funct. Mater.}, 22(17):3737�3744, 2012.  DOI: 10.1002/adfm.201200307
  
\bibitem{saville2014formation}
S.~L.~Saville, B.~ Qi, J.~ Baker, R.~ Stone, R.~E.~Camley,
  K.~L.~Livesey, L.~ Ye, T.~M. Crawford, and O.~T.~ Mefford.
\newblock The formation of linear aggregates in magnetic hyperthermia:
  Implications on specific absorption rate and magnetic anisotropy.
\newblock \emph{J. Colloid Interf. Sci.}, 424:141--151, 2014. DOI: 10.1016/j.jcis.2014.03.007

\bibitem{lammps}
S.~Plimpton. 
\newblock {Fast parallel algorithms for short-range molecular dynamics}.
\newblock \emph{J. Comp. Phys.}, 117:1-19, 1995. www.lammps.sandia.gov. DOI: 10.1006/jcph.1995.1039

\bibitem{Satoh96}
A.~Satoh, R.~W.~Chantrell, S.-I.~Kamiyama and G.~N.~Coverdale.
\newblock {Two-dimensional Monte Carlo simulations to capture thick chainlike clusters of ferromagnetic particles in colloidal dispersions}.
\newblock \emph{J. Colloid Interf. Sci.}, 178:620-627, 1996. DOI: 10.1006/jcis.1996.0159



\bibitem{pshen}A.~F.~Pshenichnikov and A.~A.~Kuznetsov.
\newblock Self-organization of magnetic moments in dipolar chains with restricted degrees of freedom.
\newblock \emph{Phys. Rev. E}, 92(4):042303, 2015. DOI: https://doi.org/10.1103/PhysRevE.92.042303

\bibitem{Faraudo16}
J.~Faraudo, J.~S.~Andreu, C.~Dalero and J.~Camacho.
\newblock {Predicting the self-assembly of superparamagnetic colloids under magnetic fields}.
\newblock \emph{Adv. Funct. Mater.}, 2016. DOI:10.1002/adfm.201504839

\bibitem{Serantes2014assemblies}
D.~ Serantes, K.~ Simeonidis, M.~ Angelakeris, O.~
  Chubykalo-Fesenko, M.~ Marciello, M.~ del~Puerto Morales, D.~
  Baldomir, and C.~ Martinez-Boubeta.
\newblock Multiplying magnetic hyperthermia response by nanoparticle
  assembling.
\newblock \emph{J. Phys. Chem. C}, 118(11):5927--5934, 2014. DOI: 10.1021/jp410717m

\bibitem{Carrey14}
R.~P.~Tan, J.~Carrey, and M.~Respaud.
\newblock Magnetic hyperthermia properties of nanoparticles inside lysosomes
  using kinetic {M}onte {C}arlo simulations: Influence of key parameters and
  dipolar interactions, and evidence for strong spatial variation of heating
  power.
\newblock \emph{Phys. Rev. B}, 90:214421, 2014. DOI: https://doi.org/10.1103/PhysRevB.90.214421

\bibitem{Ondrej}
O.~Hovorka.
\newblock Thermal activation in statistical clusters of magnetic nanoparticles.
\newblock \emph{J. Phys. D: Appl. Phys.} 50:044004, 2017. DOI:10.1088/1361-6463/aa5066


 
\bibitem{Pathria}
R.~K.~Pathria and P.~D.~Beale, \emph{Statistical Mechanics, Third Edition} (Elsevier, Oxford, 2011).

\bibitem{Landau}
L.D. Landau, E.M. Lifshitz, \emph{Fluid Mechanics. Vol. 6 (2nd ed.)} (Butterworth-Heinemann, XX, 1987). p. 65.

 
 \bibitem{Kubo}R.~Kubo. The fluctuation-dissipation theorem. \emph{Rep. Prog. Phys.} \textbf{29}, 255-284, 1966. DOI: 10.1088/0034-4885/29/1/306
 
 \bibitem{hamaker}H.~C.~Hamaker.
\newblock The London-van der Waals attraction between spherical particles.
\newblock \emph{Physica}, 4(10):1058--1072, 1937. DOI: http://dx.doi.org/10.1016/S0031-8914(37)80203-7


\bibitem{israel} J.~N.~Israelachvili.
\emph{Intermolecular and Surface Forces, Third Edition} (Elsevier, Waltham, 2011). ISBN: 978-0123919274

\bibitem{Bocquet} L.~Bocquet, J.~P.~Hansen, and J.~Piasecki. Friction tensor for a pair of Brownian particles: Spurious finite-size effects and molecular dynamics estimates. \emph{J. Stat. Phys.} 89(1):321--346, 1997. DOI: 10.1007/BF02770768

\bibitem{Leal} L.~G.~Leal. \emph{Advanced Transport Phenomena: Fluid mechanics and Convective Transport Processes} (Cambridge University Press, 2007).


 
 \bibitem{AllenBook}M.~P.~Allen and D.~J.~Tildesley. \emph{Computer Simulation of Liquids} (Clarendon Press, Oxford, 1987). ISBN: 978-0198556459
 
 \bibitem{VanGunst}W.~F.~Van Gunsteren and H.~J.~C.~Berendsen. A Leap-frog Algorithm for Stochastic Dynamics. Mol. Simul. 1:173-185, 1988. DOI: http://dx.doi.org/10.1080/08927028808080941
 
 \bibitem{Faure2011}
B.~Faure, G.~Salazar-Alvarez, and L.~Bergstrom.
\newblock Hamaker constant of iron oxide nanoparticles.
\newblock \emph{Langmuir} 27(14): 8659-8664, 2011. DOI: 10.1021/la201387d

\bibitem{Conde2008}
M.~M.~Conde, C.~Vega and A.~Patrykiejew.
\newblock The thickness of a liquid layer on the free surface of ice as obtained from computer simulation.
\newblock \emph{J. Chem. Phys.} 129:014702, 2008. DOI: 10.1063/1.2940195

 
 \bibitem{pair}J.~Garc\'ia-Otero, M.~Porto, J.~Rivas and A.~Bunde.
 \newblock Influence of dipolar interaction on magnetic properties of ultrafine ferromagnetic particles.
 \newblock  \emph{Phys. Rev. Lett.}, 84(1):167-170, 2000. DOI: 10.1103/PhysRevLett.84.167

 
 \bibitem{wozniak}M.~Wozniak, F.~R.~A.~Onofri, S.~Barbosa, J.~Yon, and J.~Mroczka.
\newblock Comparison of methods to derive morphological parameters of multi-fractal samples of particle aggregates from TEM images.
\newblock \emph{J. Aerosol Sci.}, 47:12--26, 2012. DOI: http://dx.doi.org/10.1016/j.jaerosci.2011.12.008


\bibitem{liu}J.~Liu, Z.~Wang, A.~Sheng, F.~Liu, F.~Qin, and Z.~L.~Wang.
\newblock In situ observation of hematite nanoparticle aggregates using liquid cell transmission electron microscopy. 
\newblock \emph{Environ. Sci. Technol.}, 50(11):5606-5613, 2016. DOI: 10.1021/acs.est.5b06305

\bibitem{teichmann} J.~Teichmann and K.~G.~van~den~Boogaart.
\newblock Cluster models for random particle aggregates -- morphological statistics and collison distance. 
\newblock \emph{Spat. Stat.}, 12:65-80, 2015. DOI: http://dx.doi.org/10.1016/j.spasta.2015.03.002

\bibitem{Etheridge}M.~L.~Etheridge, K.~R.~Hurley, J.~Zhang, S.~Jeon, H.~L.~Ring, C.~Hogan, C.~L.~Haynes, M.~Garwood, and J.~C.~Bischof. Accounting for biological aggregation in heating and imaging of magnetic nanoparticles. \emph{Technology} 2(3):214--228, 2014. DOI: http://dx.doi.org/10.1142/S2339547814500198




\bibitem{Berkov}
D.~V. Berkov, N.~L. Gorn, R.~Schmitz and D.~Stock.
\newblock Langevin dynamic simulations of fast remagnetization processes in ferrofluids with internal magnetic degrees of freedom.
\newblock \emph{J. Phys.-Cond. Mat.}, 18:S2595, 2006. DOI: http://dx.doi.org/10.1088/0953-8984/18/38/S05

\bibitem{Lamb}
H.~Lamb, \emph{Hydrodynamics} (Dover, New York, 1945), p.~589.

 
\bibitem{Sun2016}
S.~Sun, G.~Zhao, Y.~Huang, M.~Cai, Y.~Shan, H.~Wang and Y.~Chen.
\newblock Specificity and mechanism of action of alpha-helical membrane-active peptides interacting with model and biological membranes by single-molecule force spectroscopy.
\newblock \emph{Sci. Rep.} 6:29145, 2016. DOI: 10.1038/srep29145

\bibitem{Xue2014}
Y.~Xue, X.~Li, H.~Li, and W.~Zhang.
\newblock Quantifying thiol-gold interactions towards the efficient strength control.
\newblock \emph{Nat. Commun.} 5:4348, 2014. DOI: 10.1038/ncomms5348

\bibitem{Dong2011}
M.~Dong and O.~Sahin.
\newblock A nanomechanical interface to rapid single-molecule interactions.
\newblock \emph{Nat. Commun.} 2:247, 2011. DOI: 10.1038/ncomms1246



\bibitem{meff}
L.~Ye, T.~Pearson, Y.~Cordeau, O.~T.~Mefford, and T.~M.~Crawford.
\newblock Triggered self-assembly of magnetic nanoparticles. 
\newblock \emph{Sci. Rep.}, 26:3983-3989, 2016. DOI: 10.1038/srep23145


 \end{thebibliography}



\end{document}